# Dynamic behavior of nanobeams under axial loads: Integral elasticity modeling and size-dependent eigenfrequencies assessment

Raffaele Barretta,

Marko Čanađija,

Francesco Marotti de Sciarra,

Ante Skoblar,

Roberto Žigulić

# Dynamic behaviour of nano-beams under axial loads: integral elasticity modelling and size-dependent eigenfrequencies assessment

Raffaele Barretta[1], Marko Čanađija[2], Francesco Marotti de Sciarra[1], Ante Skoblar [*,2], Roberto Žigulić[2]

* Corresponding author. Email: ante.skoblar@riteh.hr, Tel.: +385-51-651-498; Fax.: +385-51-651-490

[1] Department of Structures for Engineering and Architecture, University of Naples Federico II,Via Claudio 21,80121 Naples, Italy
[2] University of Rijeka, Faculty of Engineering, Department of Engineering Mechanics, Vukovarska 58, 51000 Rijeka, Croatia

## Abstract

In this article, eigenfrequencies of nano-beams under axial loads are assessed by making recourse to the well-posed stress-driven nonlocal model (SDM) and strain-driven two-phase local/nonlocal formulation (NstrainG) of elasticity and Bernoulli-Euler kinematics. The developed nonlocal methodology is applicable to a wide variety of nano-engineered materials, such as carbon nanotubes, and modern small-scale beam-like devices of nanotechnological interest. Eigenfrequencies calculated using SDM, are compared with NstrainG and other pertinent results in literature obtained by other nonlocal strategies. Influence of nonlocal thermoelastic effects and initial axial force (tension and compression) on dynamic responses are analyzed and discussed. Model hardening size effects from stress-driven approach is compared to model softening size effects from strain-driven two-phase local/nonlocal approach for increasing nonlocal parameters.

1. **Introduction**

Nowadays, analysis and modelling of mechanical behavior of small-scale tubes and beams is widely investigated [1-6] in the scientific community and there is a considerable applicative interest in engineering, medicine, and electronics [7, 8]. For example, innovative materials such as carbon nanotubes (CNT), discovered by Iijima in 1992 [9], are broadly employed in nano-sensors, nano-actuators and complex Nano-Electro-Mechanical Systems (NEMS) [10, 11].

One interesting aspect of CNTs is that they can be conveniently modelled as nanobeams, which allowed fast development of models. Methodologies of continuum mechanics are commonly exploited for estimation of size effects in small-scale devices [12, 14]. It is well-known that structures at such small scales are subjected to different physical behavior than the macroscopical counterparts. More precisely, the structures exhibit size dependence, what in turn requires application of the non-local mechanics. In the last few decades, Eringen's differential law (EDM) is certainly the most popular approach for analysis of bending and vibrations of nanobeams [15-17]. According to EDM, also called strain gradient nonlocal theory, nonlocal continuum mechanics differs from the classical (local) one since stress in a point is defined from strain in the neighboring points, not just the considered point. Also, other strain gradient theories have been proposed [18-21] to capture the size effect prominent at the micro/-nanoscale. Consequently, neglecting nonlocal effects in nanostructures can cause incorrect estimation, see [22] for statical analysis of armchair carbon nanotubes.

In Eringen's theory, the stress field is calculated by convolution integral of the elastic strain field and appropriate averaging kernel [23]. For nonlocal problems defined in unbounded domains, EDM can be considered to be equivalent to Eringen's model based on the strain-driven convolution integral [15], since the constitutive boundary conditions introduced in [24] disappear for unbounded domains. However, some authors noticed [25-27] that for nonlocal problems that include bounded domains and standard kinematics boundary conditions, EDM can cause inconsistencies and paradoxes. Remedies to alleged paradoxes are described in [28, 29]. These difficulties are bypassed in reference [30] with the use of local-nonlocal theory mixture originally introduced in [31].

Recently, in an attempt to overcome the mentioned deficiency of Eringen's nonlocal theory, a stress-driven integral model was proposed [24]. The model is based on the same philosophy as the original Eringen's theory, with the difference of defining convolution integral as a function of stress field instead of elastic strain field [32, 33], giving rise to the so-called stress-driven nonlocal model (SDM). Additionally, it was shown that suitable constitutive boundary conditions have to be imposed as well. The present research also includes constitutive boundary conditions following elaborations given in the seminal work [33].

Well-known calculations for Bernoulli-Euler (BE) beams according to classical local theory, where stress at a point depends just on the elastic strain at that point, can be found in [34-40] and, specifically for beams with initial axial force for all classical boundary conditions, in [41, 42]. This serves as a starting point for nonlocal extensions [43-48]. Just as one example [43], the Eringen's differential law (EDM) approach results in the decrease of eigenfrequencies with the growth of nonlocal parameter. Also, there are repetitive conclusions in literature that nonlocal nanoscale models effectively reduce eigenfrequencies [44, 45]. Such conclusions should be carefully analyzed because they imply that nonlocal effects cause a reduction of stiffness, with the consequence that a wide class of problems of nanomechanics cannot be tackled. SDM methodology shows a growth of eigenfrequencies with the increase of the nonlocal parameter.

Majority of nanosensors, nanobeams included, operate by exploiting eigenfrequencies, so a model that includes axial preloads can be of particular interest. To this end, present research follows the method given in [49-50] to describe the flexural motion of nanobeams under the initial axial force (tension and compression). Novel stress- and strain-driven nonlocal integral approach are rearranged into nonlocal gradient forms and can be applied to problems where eigenfrequencies increase and decrease with an increase of the nonlocal parameter. Previous results [49] are now extended to account for initially present axial loading in the case of free vibrations of nonlocal nanobeams/nanotubes. Note that temperature effects can give rise to axial loading [51-54] as well, thus the model developed in the present research includes temperature variations along the length of the nanobeam.

2. **Nonlocal framework**
   2.1. **Kinematics**

The governing model of free in-plane vibrations of a nanobeam under an initial axial load (Fig. 1) considered in this paper will be based on BE kinematical assumptions. The nanobeam is assumed to vibrate in *x-y* coordinate plane. For completeness and introduction of notation, essential points of BE kinematics are summarized below.

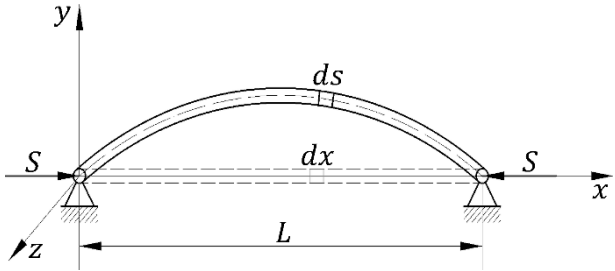

Fig. 1 A nanobeam under an initial axial load

For a nanobeam defined by BE theory, the displacement field is defined by

$$u = u_0(x,t) - yv^{(1)}(x,t), \qquad u_0(x,t) = \frac{1}{A}\int_A u(x,z,t)dA$$
$$v = v(x,t)$$
$$w = 0 \tag{1}$$

where $u(x,y,t)$ is the axial displacement of a point in a cross-section, $u_0$ represents the average displacement of a cross section, $v(x,t)$ is transverse displacement parallel to the $y$-axis in the cross-section at $x$ on the centerline ($y=0$, $z=0$) of the nanobeam, $w$ is the transverse displacement parallel to the $z$-axis, $t$ is time, Fig. 2. Superscript term in parentheses defines the order of differentiation with respect to the longitudinal coordinate $x$.

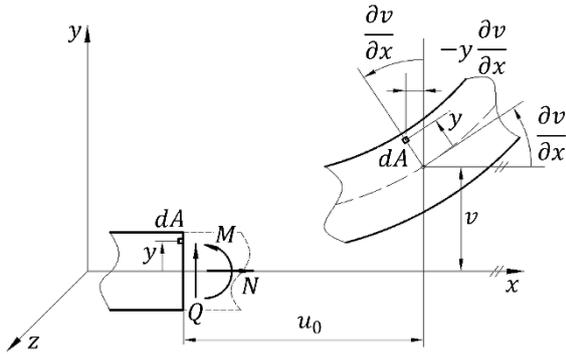

Fig. 2 Bernoulli-Euler beam theory

The strain is defined as
$$\varepsilon_{xx} = u_0^{(1)} - yv^{(2)} \tag{2}$$

It is also assumed that the strain is composed of a mechanical and thermal part:
$$\varepsilon_{xx} = \varepsilon_M + \varepsilon_T \tag{3}$$

It should be noted that the contribution of strain due to axial force is small compared to bending strain and thus neglected. Finally, the curvature is defined as:
$$\chi(x) = v^{(2)}(x). \tag{4}$$

### 2.2. Stationarity of Energy and Equilibrium

Since the focus of the present paper are free vibrations, the energy of the nanobeam does not include the external loading part related to the external bending loads. It includes the potential energy arising from bending deformations:

$$V = \frac{1}{2} \int_0^L \int_A \sigma_{xx} \varepsilon_{xx} dA dx \qquad (5)$$

and the kinetic energy related to the movement of the nanobeam:

$$T = \frac{m}{2} \int_0^L (\dot{v}^2 + \dot{u}^2) dx \qquad (6)$$

Here *m* represents the specific mass per unit beam length. The internal axial force

$$N(x,t) = \int_A \sigma_{xx} dA \qquad (7)$$

and the moment of internal forces for the *z*-axis

$$M(x,t) = -\int_A \sigma_{xx} y dA \qquad (8)$$

together with Eq. (2) provides the expression for potential energy due to bending and axial loading

$$V_b = \frac{1}{2} \int_0^L \left( N u_0^{(1)} + M v^{(2)} \right) dx \qquad (9)$$

The above results still do not include the effects of the initial axial force. The kinetic energy is usually assumed not to be influenced by the axial force [34], so the only difference between a beam with and without initial axial force is in the potential energy (7). In the bent nanobeam, the initial length of the straight centerline elongates into a bent centerline (Fig. 1) and, if the beam was initially compressed, potential energy of the beam reduces due to work of internal forces caused by the compressive external axial force *S*. For an initially tensioned nanobeam, the potential energy is increasing. During such a movement, the axial force *S* performs mechanical work on the difference of centerline lengths in straight and bent beam [36]:

$$\int_0^L (ds - dx) \sim \frac{1}{2} \int_0^L \left( v^{(1)} \right)^2 dx \qquad (10)$$

so the corresponding reduction of potential energy including the average axial displacement $u_0$ is defined by

$$V_a = S \left( \int_L \frac{1}{2} \left( v^{(1)} \right)^2 dx + u_0(0) + u_0(L) \right) \qquad (11)$$

providing the total potential energy in the presence of axial forces:

$$V = V_b + V_a = \frac{1}{2} \int_0^L M v^{(2)} dx - \frac{S}{2} \int_0^L \left( v^{(1)} \right)^2 dx + \frac{1}{2} \int_L N u_0^{(1)} dx - S u_0(0) - S u_0(L) \qquad (12)$$

Note that *S* represents the external axial force, so it takes the negative sign in tension and positive in compression.

With defined energies, the governing equation of a mathematical model is calculated from Hamilton's principle [36]

$$\int_{t_1}^{t_2} (\delta_v T - \delta_v V)\,dt = 0, \qquad \delta_v = \delta v(x,t) = 0|_{t=t_1,t_2}, \qquad 0 \leq x \leq L \tag{13}$$

which includes the variation of kinetic energy $\delta T$ and potential energy of internal forces $\delta V$ with respect to transverse displacement. Variation of kinetic energy term gives:

$$\delta_v T = \int_0^L m\dot{v}\delta\dot{v}\,dx \tag{14}$$

where superimposed dots define differentiation with respect to time, as usual.
The latter result can be integrated by parts with respect to time

$$\int_{t_1}^{t_2} \delta_v T \, dt = \int_0^L \left(m\dot{v}\delta v|_{t_1}^{t_2}\right)dx - \int_{t_1}^{t_2}\int_0^L m\ddot{v}\delta v\,dx\,dt \tag{15}$$

Using $\delta v=0$ for $t=t_1$ and $t=t_2$ eliminates the first term on the right-hand side, so the final form of the variation of kinetic energy is obtained:

$$\delta_v T = -\int_0^L m\ddot{v}\delta v\,dx \tag{16}$$

The same procedure can be repeated to obtain the variation of potential energy with respect to the transverse displacement. Starting from the first part in Eq. (11):

$$\delta_v V_b = \int_0^L M\delta v^{(2)}\,dx \tag{17}$$

and integration by parts with respect to the $x$ coordinate gives:

$$\delta_v V_b = M\delta v^{(1)}\big|_0^L - M^{(1)}\delta v\big|_0^L + \int_0^L M^{(2)}\delta v\,dx \tag{18}$$

The second part of Eq. (11) defining the influence of the axial force yields:

$$\delta_v V_a = -\int_0^L Sv^{(1)}\delta v^{(1)}\,dx \tag{19}$$

or after integration by parts:

$$\delta_v V_a = -Sv^{(1)}\delta v\big|_0^L + \int_0^L Sv^{(2)}\delta v\,dx \tag{20}$$

Inclusion of expressions for variation of kinetic Eq. (16) and potential energy Eqs. (18, 20) into Hamilton's principle (13) provides:

$$\int_{t_1}^{t_2}\left(\int_0^L \left(-m\ddot{v} - M^{(2)} - Sv^{(2)}\right)\delta v\,dx - M\delta v^{(1)}\big|_0^L + \left(M^{(1)} + Sv^{(1)}\right)\delta v\big|_0^L\right)dt = 0 \tag{21}$$

Accounting for arbitrariness of the virtual transverse displacement $\delta v$ in the first term of Eq. (21) gives the governing equation:

$$M^{(2)} + Sv^{(2)} + m\ddot{v} = 0. \tag{22}$$

Taking in arbitrariness of the virtual transverse displacement $\delta v$ and virtual rotation $\delta v^{(1)}$ at beam's ends in Eq. (21) provides the boundary conditions as:

$$\begin{aligned}\left(M^{(1)}(0) + Sv^{(1)}(0)\right)v(0) = 0, \quad -M(0)v^{(1)}(0) = 0 \\ \left(M^{(1)}(L) + Sv^{(1)}(L)\right)v(L) = 0, \quad -M(L)v^{(1)}(L) = 0\end{aligned} \tag{23}$$

The same process can be repeated for the variation with respect to axial displacement.

$$\int_{t_1}^{t_2} (\delta_u T - \delta_u V)\mathrm{d}t = 0, \quad \delta_u = \delta u_0(x,t) = 0|_{t=t_1,t_2}, \quad 0 \leq x \leq L \tag{24}$$

where

$$\delta_u T = -\int_0^L m\ddot{u}\,\delta u(x,t)dx,$$

$$\delta_u V = \int_L N\delta_0^{(1)} \mathrm{d}x - S\delta u_0(0) - S\delta u_0(L) \tag{25}$$

and after integration by parts the governing equation and boundary conditions are provided as

$$\begin{aligned} N^{(1)} &= A_\rho \ddot{u}_0, \quad A_\rho = \iint \rho(x)dA \\ N(0) &= -S, \quad N(L) = S. \end{aligned} \tag{26}$$

### 2.3. Nonlocal formulation

### 2.3.1 Stress-driven nonlocal model (SDM)

In the present formulation, the integral convolution law of elastic field using stress-driven nonlocal model is adopted [24]:

$$\varepsilon_{xx}(x,y) = \varepsilon = \int_0^L \Xi_\lambda(x-\xi)E^{-1}\sigma(\xi,y)\mathrm{d}\xi + \alpha\Delta\theta \tag{27}$$

where $E$ is Young's modulus, $\sigma$ is the axial stress, $\alpha$ is the coefficient of thermal expansion, $\Delta\theta(x,y)$ is the temperature change field, $\Xi_\lambda$ is the special convolution kernel [24]

$$\Xi_\lambda(x) = \frac{1}{2L_c}\exp\left(-\frac{|x|}{L_c}\right) \tag{28}$$

$L_c$ is the characteristic nanobeam length defined by the expression

$$L_c = \lambda L \tag{29}$$

where $L$ is the length of nanobeam and $\lambda$ is the dimensionless nonlocal parameter ($\lambda>0$).

The integral law is equivalent to the differential problem augmented with a set of constitutive boundary conditions that will be introduced later

$$\sigma = E\left(-L_c^2\left(\varepsilon^{(2)} - (\alpha\Delta\theta)^{(2)}\right) + \varepsilon - \alpha\Delta\theta\right). \tag{30}$$

With Eq. (8), the bending moment is:

$$M(x,t) = -\int_A E\left(-L_c^2\left(\varepsilon^{(2)} - (\alpha\Delta\theta)^{(2)}\right) + \varepsilon - \alpha\Delta\theta\right)y\,dA \tag{31}$$

or with Eq. (2):

$$M(x,t) = -E\left(-L_c^2\left(-I_z v^{(4)} + u_0^{(3)} S_z - (\alpha\Delta\theta)^{(2)} S_z\right) - I_z v^{(2)} + u_0^{(1)} S_z - \alpha\Delta\theta S_z\right) \tag{32}$$

where first $S_z = \int_A y\,dA$ and second moment $I_z = \int_A y^2\,dA$ of area were introduced. Considering only cross-sections for which the first moment of area vanishes ($S_z=0$), the bending moment with definition of curvature Eq. (4) is

$$M(x,t) = -L_c^2 E I_z \chi^{(2)} + E I_z \chi \tag{33}$$

or

$$\chi_{EL} - L_c^2 \chi_{EL}^2 = CM \tag{34}$$

where $C=1/K$ represents the local elastic compliance which is the inverse of elastic stiffness

$$K = \int_A E y^2 \, dA = E I_z \tag{35}$$

Replacement of the integral convolution law Eq. (31) with the gradient counterpart is only possible if the constitutive boundary conditions are enforced [29]:

$$\chi_{EL}{}^{(1)}(0,t) - \frac{1}{L_c}\chi_{EL}(0,t) = 0, \qquad \chi_{EL}{}^{(1)}(L,t) + \frac{1}{L_c}\chi_{EL}(L,t) = 0. \tag{36}$$

At the end, it should be pointed out that the temperature gradient along the nonlocal beam that are symmetric with respect to $z$ axis does not have any direct influence on the bending behavior. The main temperature contribution is by means of Eqs. (26) as the additional source of axial preload if the nanobeam dilatations are constrained.

As for the internal axial force, the identical procedure can be applied giving:

$$N(x,t) = EA\left(-L_c^2 u_0^{(3)} + L_c^2 (\alpha \Delta\theta)^{(2)} + u_0^{(1)} - \alpha \Delta\theta\right) \tag{37}$$

With the following constitutive boundary conditions:

$$\begin{aligned}
(u_0{}^{(2)}(0,t) - (\alpha\Delta\theta(0,t))^{(1)}) - \frac{1}{L_c}(u_0{}^{(1)}(0,t) - \alpha\Delta\theta(0,t)) = 0, \\
(u_0{}^{(2)}(L,t) - (\alpha\Delta\theta(L,t))^{(1)}) + \frac{1}{L_c}(u_0{}^{(1)}(L,t) - \alpha\Delta\theta(L,t)) = 0
\end{aligned} \tag{38}$$

### 2.4.2. Nonlocal strain-driven gradient model (NstrainG)

Based on [29, 50] the following class of strain-driven integral constitutive laws:

$$\sigma(x,y) = (1-\eta)\int_0^L \Xi_\lambda(x-\xi) E(\varepsilon_{xx}(\xi,y) - \alpha\Delta\theta)d\xi + \eta E(\varepsilon_{xx}(\xi,y) - \alpha\Delta\theta) \tag{39}$$

can be regarded as a special case of the nonlocal strain gradient elasticity known as the two-phase local/non-local strain-driven method. Above, $0 < \eta \leq 1$ represents a mixture parameter. Please note that the problem is well-defined for choices $\eta > 0$, while for $\eta = 0$ is ill-conditioned, see discursion in [29]. The model relies on replacement of the latter integral formulation with the gradient one, accompanied with the suitable constitutive boundary conditions. In particular, upon application of the equilibrium equations, the following relations are obtained:

$$N(x,t) = (1-\eta)EA \int_0^L \Xi_\lambda(x-\xi, L_c)(\varepsilon(\xi,t) - \alpha\Delta\theta)d\xi + EA(\varepsilon(x,t) - \alpha\Delta\theta)\eta \tag{40}$$

$$M(x,t) = (1-\eta)EI \int_0^L \Xi_\lambda(x-\xi, L_c)\chi(\xi,t)d\xi + EI\eta\chi(x,t) \tag{41}$$

It can be demonstrated that the latter integral formulation is equivalent to gradient problems:

$$\frac{1}{L_c^2}N(x,t) - N^{(2)}(x,t) = \frac{1}{L_c^2}EA(\varepsilon(x,t) - \alpha\Delta\theta) - EA\eta\varepsilon^{(2)}(x,t) \tag{42}$$

$$\frac{1}{L_c^2}M(x,t) - M^{(2)}(x,t) = \frac{1}{L_c^2}EI\chi(x,t) - EI\eta\chi^{(2)}(x,t) \tag{43}$$

equipped with suitable constitutive boundary conditions at beam ends:

$$\begin{aligned}
N^{(1)}(0,t) - \frac{1}{L_c}N(0,t) = AE(-\frac{\eta}{L_c}(\varepsilon(0,t) - \alpha\Delta\theta) + \eta(\varepsilon^{(1)}(0,t) - (\alpha\Delta\theta)^{(1)})), \\
N^{(1)}(L,t) + \frac{1}{L_c}N(L,t) = AE(\frac{\eta}{L_c}(\varepsilon(L,t) - \alpha\Delta\theta) + \eta(\varepsilon^{(1)}(L,t) - (\alpha\Delta\theta)^{(1)}))
\end{aligned} \tag{44}$$

and

$$M^{(1)}(0,t) - \frac{1}{L_c}M(0,t) = IE(-\frac{\eta}{L_c}\chi(0,t) + \eta\chi^{(1)}(0,t)), \tag{45}$$

$$M^{(1)}(L,t) + \frac{1}{L_c}M(L,t) = IE\left(\frac{\eta}{L_c}\chi(L,t) + \eta\chi^{(1)}(L,t)\right).$$

Note that standard boundary conditions used in local Bernoulli-Euler beam theory still have to be respected.

### 2.4. Solution procedure for flexural free vibrations

First of all, axial force (support reactions) should be defined. The value of axial force is affected with axial beam inertia (26) which fades with time because of damping in material, so axial acceleration $\ddot{u}_0$ is neglected when defining axial force. Thus, Eqs. (37, 40) are used only to determine the support reactions in the case of constrained expansion of the beam in a non-isothermal environment. To obtain these, as the first step, one needs to solve the differential equation (26) with SDM or NstrainG expressions for axial nonlocal formulation.

After the support reactions or external axial forces are known, the bending moment $M$, as defined in the stress-driven theory (27), is now inserted into Eq. (22):

$$\chi_{EL}^{(2)} - L_c^2 \chi_{EL}^{(4)} = C\underbrace{\left(-m\ddot{v} - Sv^{(2)}\right)}_{M^{(2)}} \tag{46}$$

and after the inclusion of expression for curvature Eq. (4), it is obtained

$$v^{(4)} - L_c^2 v^{(6)} = -Cm\ddot{v} - CSv^{(2)} \tag{47}$$

Likewise, in the two-phase local/non-local strain-driven method approach, Eq. (22) together with Eq. (43) provides

$$-\eta L_c^2 v^{(6)} + v^{(4)}(1 - L_c^2 SC) + SCv^{(2)} = L_c^2 Cm\ddot{v}^{(2)} - mC\ddot{v}. \tag{48}$$

Then, the transverse displacement is represented by means of space function $\psi(x)$, and time function $\phi(t)$ as

$$v(x,t) = \psi(x)\phi(t) \tag{49}$$

so Eq. (22) is rephrased into:

$$\psi^{(4)}(x)\phi(t) - L_c^2 \psi^{(6)}(x)\phi(t) = -Cm\psi(x)\ddot{\phi}(t) - CS\psi^{(2)}(x)\phi(t) \tag{50}$$

As for the strain-driven model, Eq. (43), the same procedure gives:

$$-\eta L_c^2 \psi^{(6)}(x)\phi(t) + \psi^{(4)}(x)\phi(t)(1 - L_c^2 SC) + SC\psi^{(2)}(x)\phi(t) = L_c^2 C\psi^{(4)}(x)\phi^{(2)}(t) - Cm\psi(x)\ddot{\phi}(t). \tag{51}$$

It is assumed that the specific form of the time function, $\phi(t)$, is in correlation with its second derivation with respect to time as:

$$\ddot{\phi}(t) + \omega^2 \phi(t) = 0 \tag{52}$$

where $\omega$ is the eigenfrequency of the nanobeam in the case of free vibrations. The solution of Eq. (52) implies that the time function has the form:

$$\phi(t) = a\sin(\omega t) + b\cos(\omega t) \tag{53}$$

Eq. (52) can be introduced into Eq. (50), what eventually provides the space function (i.e. mode shape) for the stress-driven model:

$$\psi^{(4)}(x) - L_c^2 \psi^{(6)}(x) + CS\psi^{(2)}(x) - Cm\omega^2 \psi(x) = 0 \tag{54}$$

which is a homogeneous differential equation of sixth order with constant coefficients and the general solution:

$$\psi(x) = \sum_{i=1}^{6} C_i \psi_i(x) \tag{55}$$

Functions $\psi_i(x)$ are linearly independent solutions of Eq. (54) and $C_i$ are unknown constants.
Finally, application of six boundary conditions, Eq. (23, 36) on the general solution (55) gives six homogenous algebraic equations with six unknowns $C_i$ and unknown eigenfrequency $\omega$.

Similarly, the strain-driven counterpart is:

$$-\eta L_c^2\psi^{(6)}(x) + \psi^{(4)}(x)(1 - L_c^2 SC) + SC\psi^{(2)}(x) = -\omega^2(L_c^2 C\psi^{(4)}(x) - Cm\psi(x)). \quad (56)$$

with boundary conditions (23, 45).

The solution procedure can be described as follows:
- Algebraic equations can be rewritten in the matrix form as the product of a quadratic matrix of linearly independent equation solutions, $\mathbf{A}(\lambda,\omega)$, and the vector of constants $C_i$, $\mathbf{p}$:
$$\mathbf{A}(\lambda, \omega)\mathbf{p} = 0 \quad (57)$$
- If the determinant of matrix $\mathbf{A}(\lambda,\omega)$ is equal to zero, then the system has either no nontrivial solutions or an infinite number of solutions. Thus, the values of eigenfrequencies $\omega$ are defined as zero points of the determinant
$$\det \mathbf{A}(\lambda, \omega) = 0 \quad (58)$$
for chosen dimensionless nonlocal parameter $\lambda$.
- The zero points are obtained numerically.

## 3. Examples

### 3.1. Eigenfrequencies of a simply supported nanobeam

In this example, the influence of the increase of dimensionless nonlocal parameter $(0 \leq \lambda \leq 0.1)$ on the first dimensionless eigenfrequency of a simply supported beam for different cases of discrete dimensionless initial axial forces ($S^*=-1$, $S^*=0$ and $S^*=1$) will be analyzed. Subsequently, a more detailed analysis will be performed for $\lambda=0.1$. Solutions in this and other examples were obtained by the aid of Wolfram Mathematica software.

Dimensionless eigenfrequency, $\omega^*$, and dimensionless initial axial force, $S^*$, are defined with expressions:
$$\omega^* = \omega_1 L^2 \sqrt{\frac{m}{K}} \quad \text{and} \quad S^* = \frac{SL^2}{EI} \quad (59)$$

For a simply supported nanobeam on both ends, the classical boundary conditions are valid:
$$\begin{array}{ll} v(0) = 0, & M(0) = 0 \\ v(L) = 0, & M(L) = 0 \end{array} \quad (60)$$

Two constitutive boundary conditions Eq. (36) must be also fulfilled. Consequently, six conditions are available.

Solutions are obtained by following the procedure described in Sec. 2.4. Calculated eigenfrequencies are compared to published results for a nanobeam without dimensionless initial axial force, $S^*=0$, also calculated with SDM theory [49] and to results obtained with the Stress theory with dimensionless Effective nonlocal Moment (SEM) [48] (Fig. 3). Dimensionless effective nonlocal bending moment is defined as

$$\bar{M}_{ef} = \bar{M} - 2\sum_{n=1}^{\infty} \lambda^{2n}\bar{M}^{\langle 2n\rangle} \quad (61)$$

where $\bar{M}$ is the dimensionless nonlocal bending moment of nanobeam defined as
$$\bar{M} = \frac{ML}{EI_z} \quad (62)$$

Where $M$ is local bending moment. The procedure involves defining governing equation as the sixth-order partial differential equation and six nonclassical higher order boundary conditions [48, 55].

The values of calculated dimensionless eigenfrequencies are also listed in Table 1. SDM theory gives lower eigenfrequency than SEM but both theories give hardening size effects to the model while NstrainG theory gives softening size effects. Loading with compressive forces (*S*>1) also lowers the first eigenfrequency. Tension loading has the opposite effect (Fig. 4).

The first point at the left-hand side of diagrams (*λ*=0) for three initial axial forces is equal to eigenfrequency calculated with local theory without the influence of nonlocal parameter [41, 42].

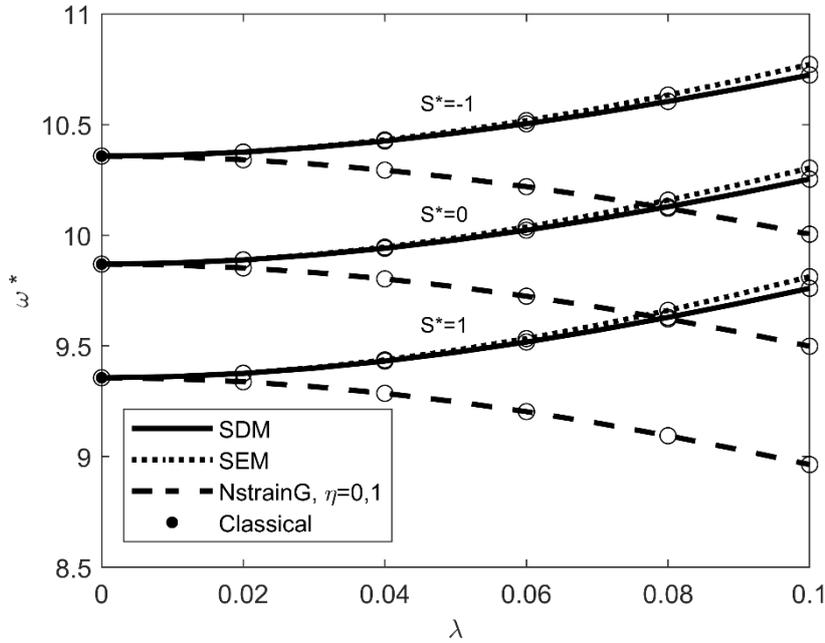

Fig. 3 First dimensionless eigenfrequency of the simply supported nanobeam as a function of nonlocal parameter and initial axial force

The same nanobeam will be analyzed in more detail for the constant dimensionless nonlocal parameter *λ*=0.1. Diagram in Fig. 4 compares results obtained by SDM and NstrainG theory to SEM theory [48] and classical (local) theory [41]. Same results are also given in Table 2.

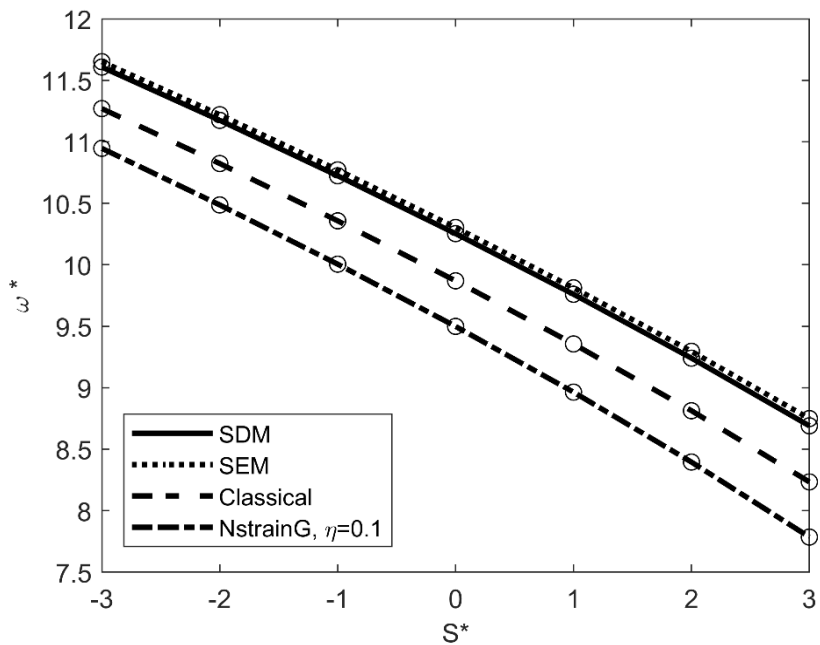

Fig. 4 First dimensionless eigenfrequency of the simply supported nanobeam as a function of initial axial force for $\lambda$=0.1

### 3.2. Influence of nanobeam length on eigenfrequencies

This example investigates the influence of the nanobeam length on eigenfrequency. The nanobeam is simply supported with characteristic length $L_c$=2 nm, cross-section height 4 nm and width 10 nm, density $\rho$=2400 kg/m$^3$, initial axial tension force 1 nN and modulus of elasticity $E$=100 GPa.

Classical boundary conditions for a simply supported beam are the same as in Example 3.1. Solutions are again obtained by following the procedure introduced in Sec. 2.4. Fig. 5 and Table 3 compares the results of SDM, NstrainG, SEM and classical (local) theory. The increase in the length ($L$ =70, 80, 90, 100, 110 and 120 nm) of the nanobeam results in the decrease of the nonlocality, converging to the classical (local) solution for longest nanobeams.

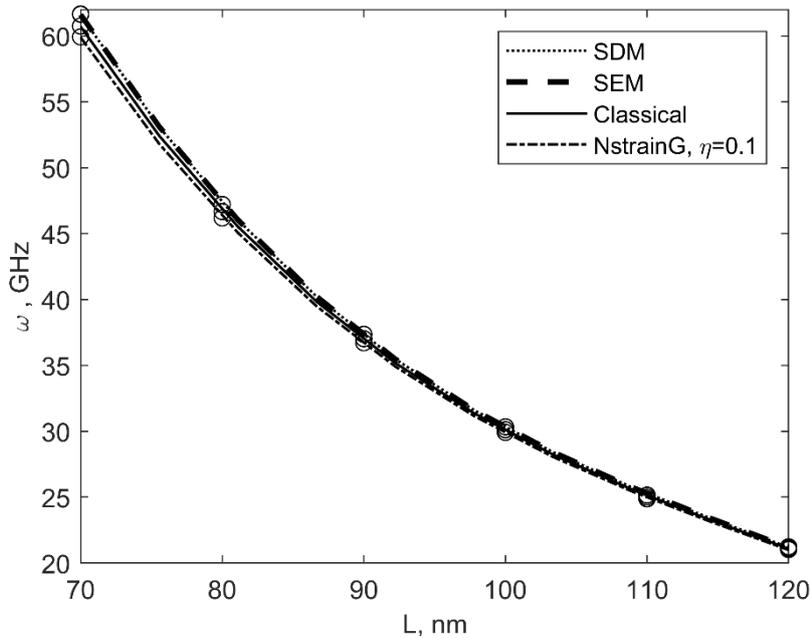

Fig. 5 Second eigenfrequency of the simply supported nanobeam as a function of the nanobeam length with fixed characteristic length

### 3.3. Nonlocal cantilever nanotube

In this example, the influence of nonlocal parameter ($\lambda$=0$^+$-0.1) on the first eigenvalue of a cantilever nanotube with different initial axial forces ($S^*$=-2, -1, 0, 1 and 2) will be analyzed. The nanotube is conveniently modeled as a nanobeam, so the herein proposed formulation can be applied.

In addition to the constitutive boundary conditions Eq. (36), for the cantilever nanobeam the classical boundary conditions must be respected as well:

$$v(0) = 0, \quad v^{(1)}(0) = 0$$
$$M^{(1)}(L) + Sv^{(1)}(L) = 0, \quad M(L) = 0 \tag{63}$$

Fig. 6 and Table 4 present results of SDM and NstrainG theory compared to the classical theory [41, 42] and to published results of SDM theory for nano-beam without initial axial stress $S^*=0$ [49]. The trend noticed in the case of the simply supported nanobeam is visible in the present case as well: increase of the tension axial force will increase the stiffness of the nanotube, giving higher eigenfrequencies. On the other hand, the increase of the nonlocal parameter $\lambda$ raises the first eigenfrequency for SDM and lovers it for NstrainG theory.

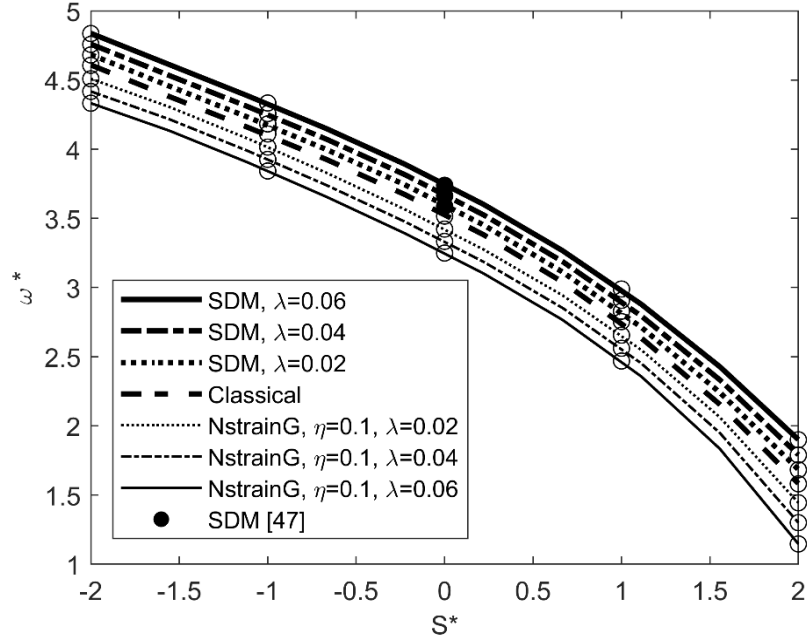

Fig. 6 First dimensionless eigenfrequency of the cantilever nanobeam as a function of initial axial force and nonlocal parameter

### 3.4. Doubly-clamped nanobeam in linear temperature field

The final example illustrates influence of thermal effects on the nanobeam eigenfrequencies. In particular, a temperature field that linearly varies in the longitudinal direction $\Delta\theta = \theta_0 x$ is considered. The nanobeam has a characteristic length $L_c$=5 nm, cross-section height 4 nm and width 10 nm, density $\rho$=2400 kg/m$^3$, modulus of elasticity $E$=100 GPa, coefficient of thermal expansion $\alpha$=0.1, mixture parameter $\eta$=0.1 and dimensionless nonlocal parameter $\lambda$=0.01.

Since the nanobeam is clamped at both ends, the expansion due to temperature is not possible, consequently leading toward axial support reactions at both ends. Solution will be obtained by the strain-driven formulation.

Axial loads follow from Eq. (26). The solution is obtained from Eq. (42) accompanied with classical and constitutive boundary conditions Eqs. (44):

$$\frac{1}{L_c^2} N(x,t) - N^{(2)}(x,t) - \frac{1}{L_c^2} EA\left(u_0^{(1)}(x,t) - \alpha\theta_0 x\right) + EA\eta u_0^{(3)}(x,t) = 0, \tag{64}$$

equipped with suitable constitutive boundary conditions at beam ends:

$$\begin{aligned}
N^{(1)}(0,t) - \frac{1}{L_c} N(0,t) &= AE\left(-\frac{\eta}{L_c}(u_0^{(1)}(0,t) - \alpha\Delta\theta) + \eta(u_0^{(2)}(0,t) - (\alpha\Delta\theta)^{(1)})\right), \\
N^{(1)}(L,t) + \frac{1}{L_c} N(L,t) &= AE\left(\frac{\eta}{L_c}(u_0^{(1)}(L,t) - \alpha\Delta\theta) + \eta(u_0^{(2)}(L,t) - (\alpha\Delta\theta)^{(1)})\right)
\end{aligned} \tag{65}$$

where $N^{(1)} = 0$ since beam axial acceleration $\ddot{u}_0$ is neglected in definition of the axial force.

The above procedure, with $u_0(0) = u_0(L) = 0$, gives the axial support force at beam ends as:

$$S = \frac{AE\alpha\theta_0 L \left(-1 + e^{1/(\lambda\sqrt{\eta})}(1+\sqrt{\eta}) + \sqrt{\eta}\right)}{2 - 2\sqrt{\eta} - 4\eta\lambda - 2e^{1/(\lambda\sqrt{\eta})}(1 + \sqrt{\eta} - 2\lambda\eta)}.$$

Now, influence of the preload on the eigenfrequencies can be estimated by following the procedure illustrated in previous examples.

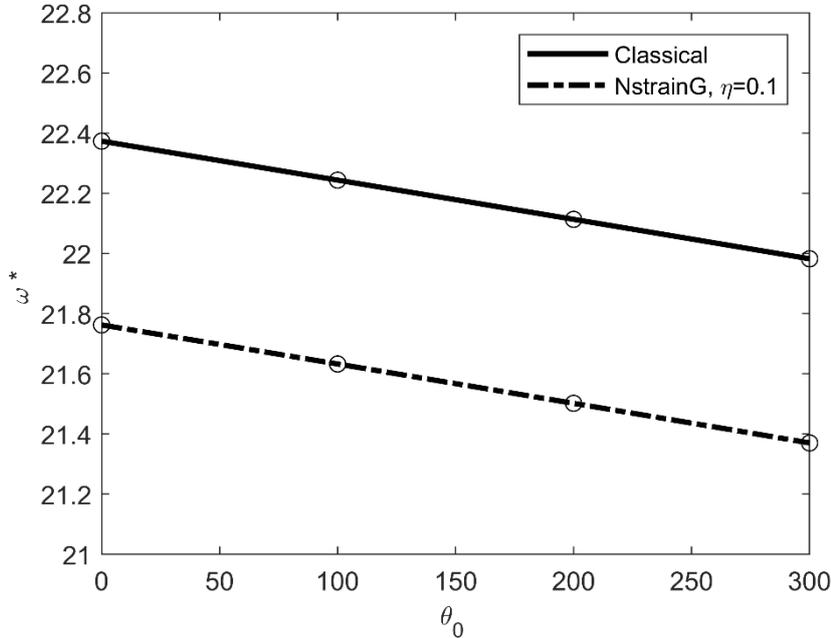

Fig. 7 First dimensionless eigenfrequency of the clamped nanobeam as a function of temperature field coefficient $\theta_0$

Fig. 7 and Tab. 5 clearly demonstrate that temperature increase will also increase forces in supports, what will lead towards reduction of the first eigenfrequency.

## 4. Conclusions

It is well-known from the classical local theory of elasticity that eigenfrequencies of beams change under the action of axial loading. The effect is present in nanostructures as well. However, since the nanostructures are highly influenced by size-effects, nonlocal methodologies should be used instead of classical local formulations. In the line with this, the paper at hand extended the local BE solution for calculating eigenfrequencies to nonlocal regime suitable for nanobeams/nanotubes that are initially axially loaded. A consistent stress- and strain-driven nonlocal integral model and Bernoulli-Euler's kinematics were rewritten in the nonlocal gradient setting, thus paradoxes exhibited by some other approaches are avoided.

Detailed analyses provided in the Example section give rise to the following main conclusions:
- Nonlocal effects and initial axial force (either of mechanical or thermal origin) on a nanobeam cause a noticeable change of eigenfrequencies.
- An increase of initial axial tension force causes the increase of eigenfrequencies while an increase of the initial axial compression force causes a reduction of eigenvalues for a fixed nonlocal parameter. Similar behavior is also seen in classical local theory.
- Independently of the tensile or compressive nature of the axial force, increasing nonlocal parameters cause the increase of eigenfrequencies implicating higher nanobeam stiffness for stress-driven method and the decrease of eigenfrequencies implicating lower nanobeam stiffness for strain-driven method.
- Numerical results show convergence of eigenvalues to the classical ones with reduction of the nonlocal parameter to zero value or with the growth of nanobeam geometric dimensions.
- Presented stress- and strain-driven formulations can be applied to nanotechnology, sensors in particular. Since the model accounts for the axial preloads and longitudinal temperature effects, a more general application field can be expected.

**Acknowledgments**
This work has been partially supported by Croatian Science Foundation under the project IP-2019-04-4703, partially by the University of Rijeka under the project number uniri-tehnic-18-37, partially by the University of Rijeka under the project number uniri-tehnic-18-225, and partially by the Italian Ministry for University and Research P.R.I.N. (National Grant 2017, Project code 2017J4EAYB "University of Naples Federico II" Research Unit). This support is gratefully acknowledged.

Table 1 First dimensionless eigenfrequency of a simply supported nanobeam

| $\omega^*$ | $\lambda$ | | | | | |
|---|---|---|---|---|---|---|
| | 0 | 0.02 | 0.04 | 0.06 | 0.08 | 0.1 |
| $S^*=0$ | | | | | | |
| SDM | 9.869604 | 9.888289 | 9.941046 | 10.022791 | 10.128486 | 10.253421 |
| SEM | 9.869604 | 9.888991 | 9.946024 | 10.037497 | 10.158564 | 10.303367 |
| NstrainG, $\eta=0.1$ | 9.869604 | 9.852459 | 9.802950 | 9.724650 | 9.621770 | 9.498803 |
| Classical | 9.869604 | | | | | |
| $S^*=1$ | | | | | | |
| SDM | 9.356254 | 9.375962 | 9.431583 | 9.517696 | 9.628915 | 9.760204 |
| SEM | 9.356254 | 9.376702 | 9.436832 | 9.533191 | 9.660581 | 9.812735 |
| NstrainG, $\eta=0.1$ | 9.356254 | 9.338166 | 9.285913 | 9.203208 | 9.094404 | 8.964146 |
| Classical | 9.356254 | | | | | |
| $S^*=-1$ | | | | | | |
| SDM | 10.357543 | 10.375349 | 10.425643 | 10.503624 | 10.604549 | 10.723977 |
| SEM | 10.357543 | 10.376017 | 10.430389 | 10.517649 | 10.633251 | 10.771674 |
| NstrainG, $\eta=0.1$ | 10.357543 | 10.341207 | 10.294049 | 10.219520 | 10.121694 | 10.004928 |
| Classical | 10.357543 | | | | | |

Table 2 First dimensionless eigenfrequency for different axial forces and $\lambda=0.1$

| $\omega^*$ | $S^*$ | | | | | | |
|---|---|---|---|---|---|---|---|
| | -3 | -2 | -1 | 0 | 1 | 2 | 3 |
| SDM | 11.608003 | 11.174736 | 10.723974 | 10.253421 | 9.760204 | 9.240699 | 8.690192 |
| SEM | 11.651960 | 11.220453 | 10.771674 | 10.303367 | 9.812734 | 9.296244 | 8.749317 |
| Classical | 11.270222 | 10.823507 | 10.357543 | 9.869604 | 9.356254 | 8.813058 | 8.234092 |
| NstrainG, $\eta=0.1$ | 10.947199 | 10.486653 | 10.004928 | 9.498803 | 8.964146 | 8.395506 | 7.785443 |

Table 3 First and second eigenfrequencies as functions of the nanobeam length

| ω, GHz | $L$, nm | | | | | |
|---|---|---|---|---|---|---|
|  | 70 | 80 | 90 | 100 | 110 | 120 |
| SDM | 61.642 | 47.199 | 37.351 | 30.333 | 25.156 | 21.226 |
| SEM | 61.667 | 47.213 | 37.359 | 30.339 | 25.159 | 21.229 |
| Classical | 60.747 | 46.671 | 37.020 | 30.116 | 25.008 | 21.122 |
| NstrainG, η=0.1 | 59.929 | 46.189 | 36.718 | 29.918 | 24.873 | 21.026 |

Table 4. First dimensionless eigenfrequency for the cantilever nanobeam

| $\omega^*$ | S* | | | | |
|---|---|---|---|---|---|
| | -2 | -1 | 0 | 1 | 2 |
| Classical | 4.606570 | 4.110242 | 3.516015 | 2.753624 | 1.580913 |
| $\lambda=0.02$ | | | | | |
| SDM | 4.681316 | 4.182819 | 3.587733 | 2.828541 | 1.682044 |
| SDM [49] | | | 3.587734 | | |
| NstrainG | 4.509013 | 4.015483 | 3.422333 | 2.655503 | 1.443864 |
| $\lambda=0.04$ | | | | | |
| SDM | 4.757927 | 4.257533 | 3.662122 | 2.907210 | 1.788481 |
| SDM [49] | | | 3.662122 | | |
| NstrainG | 4.417778 | 3.926411 | 3.333484 | 2.560599 | 1.299844 |
| $\lambda=0.06$ | | | | | |
| SDM | 4.836092 | 4.334099 | 3.738905 | 2.989257 | 1.899160 |
| SDM [49] | | | 3.738905 | | |
| NstrainG | 4.33249 | 3.842718 | 3.249259 | 2.46882 | 1.146067 |

Table 5. First dimensionless eigenfrequency for the doubly-clamped nanobeam

| $\omega^*$ | $\theta$ | | | |
|---|---|---|---|---|
| | 0 | 100 | 200 | 300 |
| Classical | 22.373285 | 22.243800 | 22.113479 | 21.982310 |
| NstrainG, $\eta=0.1$ | 21.762548 | 21.632660 | 21.501908 | 21.370278 |